\documentclass{iau}

\usepackage{amsmath}
\usepackage{graphicx}
\usepackage{multirow}

\usepackage{gensymb}
\usepackage{enumerate}
\newcommand{\kms}{\ifmmode{\,\rm{km}\, \rm{s}^{-1}}\else{$\,$km$\,$s$^{-1}$}\fi}
\newcommand{\msun}{\ifmmode{~\rm{M}_{\odot}}\else{M$_{\odot}$}\fi}
\newcommand{\mstar}{\ifmmode{M_{\star}}\else{$M_{\star}$}\fi}

\newcommand{\logm}{\ifmmode{\log(M_{\star}/M_{\odot})}\else{$\log(M_{\star}/M_{\odot})$}\fi}
\newcommand{\mm}{\ifmmode{M_{\star}/M_{\odot}}\else{$M_{\star}/M_{\odot}$}\fi}

\newcommand{\gaia}{\textit{Gaia}}

\newcommand\aap{A\&A}                % Astronomy and Astrophysics
\newcommand\aapr{A\&ARv}             % Astronomy and Astrophysics Review (the)
\newcommand\aj{AJ}                   % Astronomical Journal (the)
\newcommand\apj{ApJ}                 % Astrophysical Journal
\newcommand\apjl{ApJ}                % Astrophysical Journal, Letters
\newcommand\araa{ARA\&A}             % Annual Review of Astronomy and Astrophysics
\newcommand\mnras{MNRAS}             % Monthly Notices of the Royal Astronomical Society
\newcommand\nat{Nature}              % Nature
\newcommand\pasp{PASP}               % Publications of the Astronomical Society of the Pacific
\newcommand\pasa{PASA}     % Publications of the Astronomical Society of Australia

\begin{document}

\lefttitle{J. van de Sande \& the GECKOS Survey collaboration}
\righttitle{The GECKOS Survey}

\jnlPage{1}{7}
\jnlDoiYr{2021}
\doival{10.1017/xxxxx}

\aopheadtitle{Proceedings IAU Symposium}
\editors{F. Tabatabaei,  B. Barbuy \&  Y. Ting, eds.}

%\title[The GECKOS Survey] %% give here short title %%
\title{GECKOS: Turning galaxy evolution on its side with deep observations of edge-on galaxies}

%\author[van de Sande et al.]   %% give here short author list %%
\author{J. van de Sande$^{1,2}$,
A. Fraser-McKelvie$^{3,2}$, \\
D. B. Fisher$^{4,2}$,
M. Martig$^{5}$,
M. R. Hayden$^{1,2}$, \\
and the GECKOS Survey collaboration}

\affiliation{$^1${Sydney Institute for Astronomy, School of Physics, A28, The University of Sydney, NSW, 2006, Australia. Email: {\tt jesse.vandesande@sydney.edu.au}} \\
$^2${ARC Centre of Excellence for All Sky Astrophysics in 3 Dimensions (ASTRO 3D), Australia}\\
$^3${International Centre for Radio Astronomy Research, The University of Western Australia, 35 Stirling Highway, Crawley WA 6009, Australia}\\
$^4${Centre for Astrophysics and Supercomputing, Swinburne University of Technology, PO Box 218, Hawthorn, VIC 3122, Australia}\\
$^5${Astrophysics Research Institute, Liverpool John Moores University, 146 Brownlow Hill, Liverpool L3 5RF, UK}}

%\pubyear{2023}
\volno{377}  %% insert here IAU Symposium No.
%\setcounter{page}{1}
%\jname{Early Disk-Galaxy Formation - from JWST to the Milky Way}
%\editors{F. Tabatabaei, B. Barbuy \& Y. Ting, eds.}
%\begin{document}

%\maketitle

\begin{abstract}
We present GECKOS (Generalising Edge-on galaxies and their Chemical bimodalities, Kinematics, and Outflows out to Solar environments), a new ESO VLT/MUSE large program. The main aim of GECKOS is to reveal the variation in key physical processes of disk formation by connecting Galactic Archaeology with integral field spectroscopic observations of nearby galaxies. Edge-on galaxies are ideal for this task: they allow us to disentangle the assembly history imprinted in thick disks and provide the greatest insights into outflows. The GECKOS sample of 35 nearby edge-on disk galaxies is designed to trace the assembly histories and properties of galaxies across a large range of star formation rates, bulge-to-total ratios, and boxy and non-boxy bulges. GECKOS will deliver spatially resolved measurements of stellar abundances, ages, and kinematics, as well as ionised gas metallicities, ionisation parameters, pressure, and inflow and outflow kinematics; all key parameters for building a complete chemodynamical picture of disk galaxies. With these data, we aim to extend Galactic analysis methods to the wider galaxy population, reaping the benefits of detailed Milky Way studies, while probing the diverse mechanisms of galaxy evolution.
\end{abstract}


\maketitle
%
%\keywords{Galaxy: abundances -- Galaxy: kinematics and dynamics -- Galaxy: stellar content -- Galaxy: structure --  Galaxy: evolution --galaxies: abundances -- galaxies: kinematics and dynamics  -- galaxies: stellar content -- galaxies: structure -- galaxies: evolution}
%\end{abstract}

%\firstsection % if your document starts with a section,
              % remove some space above using this command.
\section{Introduction}

\noindent Due to the complexity of internal and external processes acting on disk galaxies, many open questions regarding their evolution remain. Because of our unique vantage point, the Milky Way is by far the best studied galaxy in the Universe and an ideal laboratory for testing our theories (Bland-Hawthorn \& Gerhard 2016).
The ESA-\gaia{} mission (Gaia Collaboration et al. 2021) and Galactic Archaeology surveys (e.g.~APOGEE- Majewski et al. 2017, GALAH- Buder et al. 2021, LAMOST/LEGUE- Deng et al. 2012), are currently building a complex picture, where the Galaxy is shaped through {minor accretion events} and the {delicate interplay} of chemical and dynamical processes (e.g. Xiang \& Rix 2022). 

One of the most interesting recent developments is the recognition that the Galactic structural thick disk is distinct from the dominant thin disk through its unique chemistry (Bensby et al. 2014, Masseron \& Gilmore 2015, Hayden et al. 2015). In addition to its older age and higher elevation about the plane, the thick disk is found to be enhanced in [$\alpha$/Fe] as compared to the thin disk over a wide range in [Fe/H]. However, there is no consensus on how the thin/thick and $\alpha$-poor/rich disks formed and evolved. 

The distinct chemical bimodality of stars in the solar neighbourhood could either indicate a bimodal formation history (Chiappini et al. 1997, Haywood et al. 2016), or a natural consequence of how the $\alpha$-elements are produced (Sch{\"o}nrich \& Binney 2009, Vincenzo \& Kobayashi 2020, Sharma et al. 2021). The origin of the structural thick disk in the Milky Way could also be explained via some combination of (1) radial migration and slow vertical heating (Minchev et al. 2015), (2) violent disk instabilities in clumpy gas-rich disks at high redshift (Clarke et al. 2019), or (3) a series of mergers (Renaud et al. 2021). Hence, the chemodynamic history of every disk appears to be a complex mix of \emph{internal} and \emph{external} processes, the combination of which varies from galaxy to galaxy in a stochastic manner, and with too many free variables for only one target to resolve. A promising path to meet this challenge is to expand our high-resolution studies of galaxies beyond the Milky Way.

Currently, significant progress is being made on resolving the small-scale physics driving galaxy evolution via small samples of highly-spatially-resolved nearby galaxies. Several MUSE integral field spectroscopic (IFS) surveys of face-on galaxies such as PHANGS (Emsellem et al. 2022), MAD (den Brok et al. 2020), and TIMER (Gadotti et al. 2019) study disks on scales of 50-100~pc. While small in sample size, these surveys probe a spatial resolution and observational depth beyond the reach of large IFS surveys such as CALIFA (S{\'a}nchez et al. 2012), SAMI (Croom et al. 2012), and MaNGA (Bundy et al. 2015). While face-on galaxies are excellent for studying the interstellar medium and the thin disk, most of the evolutionary history is encoded in the vertical structure, which is lost due to the degeneracy with the brighter young stars in the thin disk as well as with deprojecting vertical height. 

Deep observations of edge-on systems reveal insights into mechanisms not probed otherwise. The power in studying edge-on galaxies comes from the ease in separating the vertical distributions of stars and gas. Given that the fraction of accreted stars increases with both galactocentric radius and height above the midplane, a galaxy's merger history is best extricated from these off-plane regions (e.g.~Martig et al. 2021). While structural thick disks have been shown to be a common feature in external galaxies (Yoachim \& Dalcanton 2006, Comer{\'o}n et al. 2018, Mart{\'\i}nez-Lombilla \& Knapen 2019), detecting chemically enriched disks has been more challenging (Yoachim \& Dalcanton 2008). Indeed, only recently have  $\alpha$-enhanced disks been discovered in external disk galaxies (Eigenbrot \& Bershady 2018, Pinna et al. 2019a,b, Scott et al. 2021, Martig et al. 2021).

Unique insight into galactic outflows also comes from studying nearby, edge-on systems (e.g. M82 - Shopbell \& Bland-Hawthorn 1998, NGC253 - Bolatto et al. 2013) where the substructure of outflowing mass is resolved without systematic uncertainties introduced from decomposing the disk and outflow emission. MUSE observations have already captured the spatial distribution and physical drivers of ionisation in one outflow (Bik et al. 2018). Studies of kinematics (Krieger et al. 2019) and density (Lopez et al. 2020) in two prototypical edge-on starbursts (NGC~253 \& M82) suggest that the outflow model used for the past 30~years (adiabatically cooling winds) fails. 

Furthermore, faint edge-on structures such as boxy-peanut bulges (the edge-on projection of a buckled stellar bar) are only observable in highly-inclined galaxies (e.g. Kuijken \& Merrifield 1995), and can provide a unique view on the radial migration history of discs (e.g. Martinez-Valpuesta \& Gerhard 2013). However, no systematic study of edge-on galaxies with deep IFS data yet exists; the few existing studies focus on single objects, or strictly target the inner regions. Hence, there is a clear opportunity for a comprehensive study on a representative sample of edge-on galaxies in the nearby Universe.

\begin{figure}[t]
    \includegraphics[height=3.89cm]{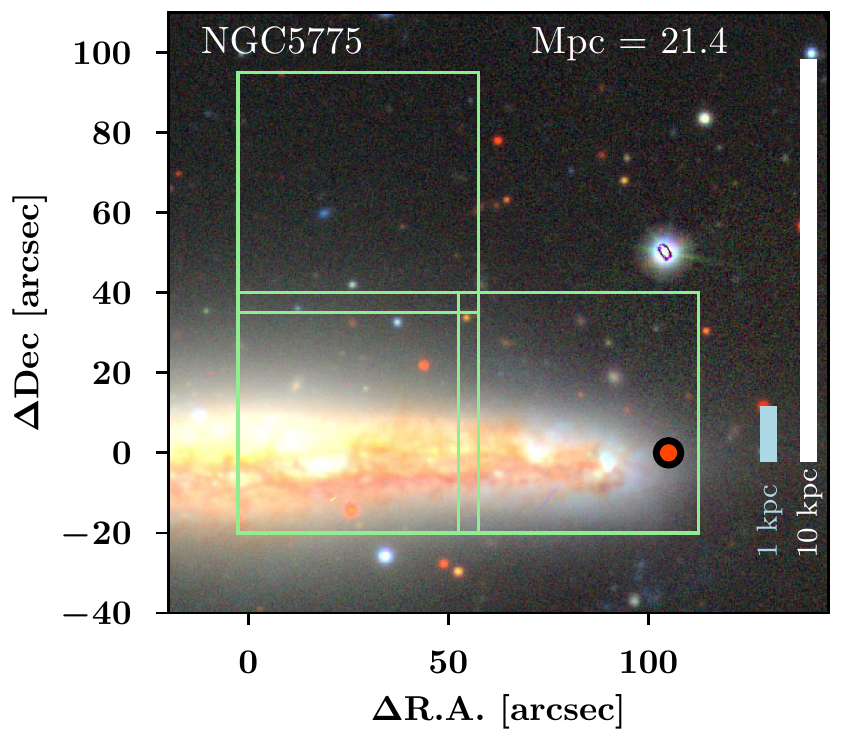}
    \includegraphics[height=3.89cm]{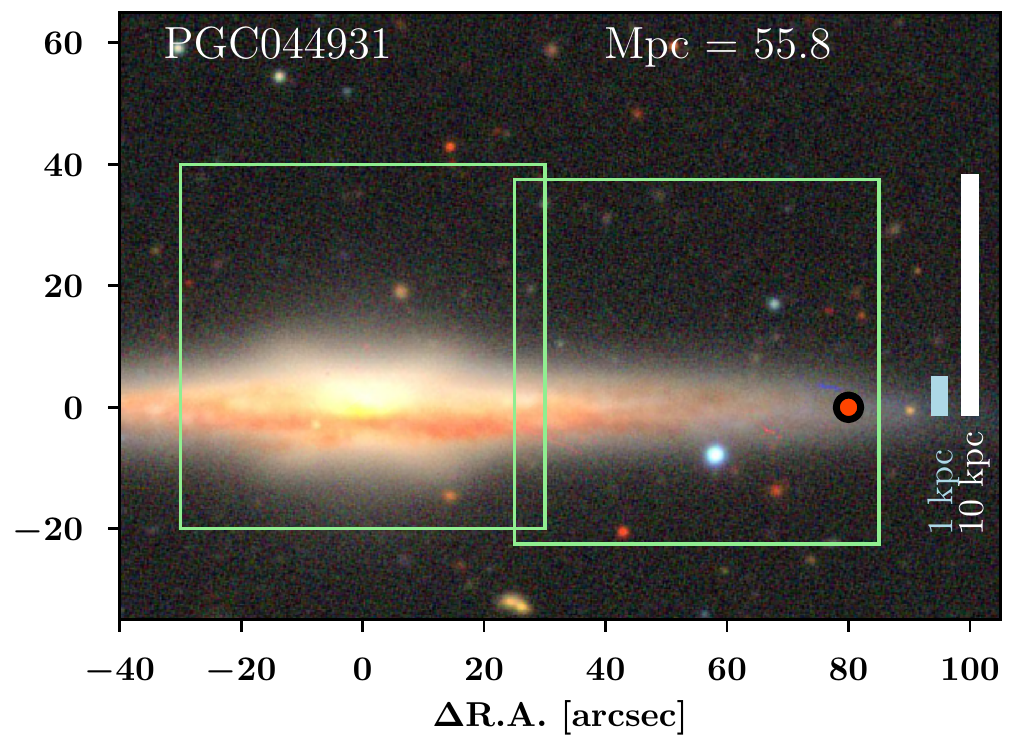}
    \includegraphics[height=3.89cm]{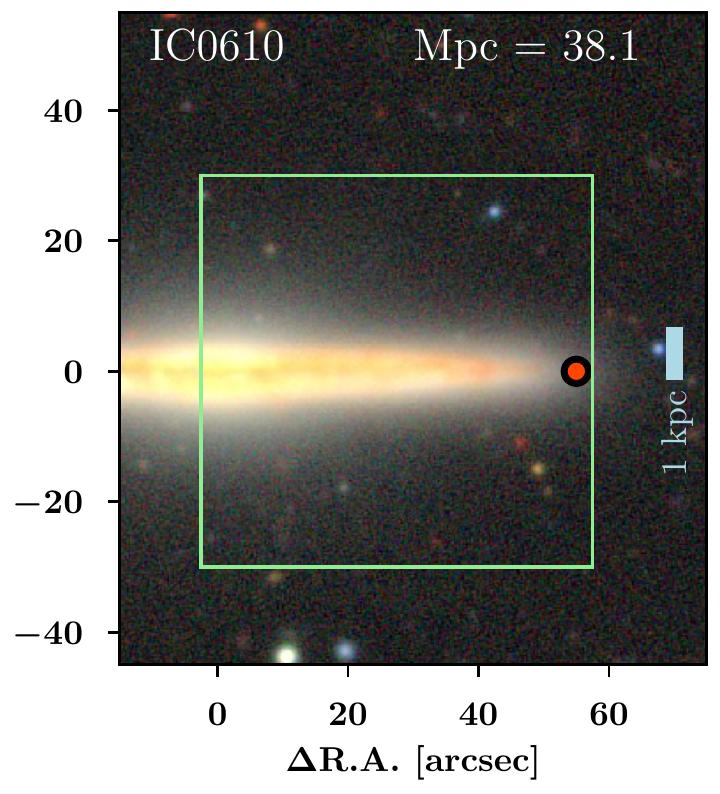}
    \caption{Three GECKOS galaxies selected from above the star-forming (SF) main sequence  (left), on the SF sequence (middle), and below the SF sequence (right). Depending on the distance, multiple MUSE pointings (green squares) are required to determine the chemodynamical properties of the stars out to the solar-like environments at $\mu_g$=$23.5$~mag~arcsec$^{-2}$ (red-black circle) and to measure outflows 10~kpc off the plane in main-sequence and high-SFR galaxies.
    } \label{fig:geckos_sample}
\end{figure}

%\vspace{-0.5cm}
\section{The GECKOS Survey}
\noindent GECKOS (Generalising Edge-on galaxies and their Chemical bimodalities, Kinematics, and Outflows out to Solar environments) is a new survey to systematically study 35 edge-on disk galaxies with the MUSE IFS, going out to larger radius, deeper (S/N${>40}$ at $\mu_g=23.5~$mag/arcsec$^{2}$), and with higher spatial resolution ($<$200 pc) than existing IFS surveys. The survey's science goals are focused on answering outstanding questions around three major themes:
\vspace{0.05cm}

\begin{enumerate}[1.]
\item~ \textbf{History and Impact of External Processes:} What is the effect of mergers on bulges and on the vertical structure of disks? Can the chemical bimodality in galaxies be explained through accretion alone?

\item~ \textbf{History and Impact of Internal Processes:} Are thick disks the remnants of a turbulent, clumpy, gas-rich phase in galaxies at $z\approx2$? How important is radial migration for thick disks? What is the role of outflows and Galactic fountains in altering galaxy chemistry?

\item~ \textbf{Synthesis - Chemodynamical Models:} Can the Milky Way be used as a template for disk evolution? What is the predictive power of Galactic chemical evolution models to explain the properties of $L*$ disk galaxies?
\end{enumerate}

\noindent {\bf The GECKOS Sample:}
The survey's aim is to capture the full range of physical mechanisms that shape the chemodynamic evolution of disk galaxies. Targets are selected from the S4G survey (66\%, Sheth et al. 2010) and HyperLEDA (34\%, Makarov et al. 2014)
within a distance range of $15<D$ [Mpc]$<70$. We specifically target edge-on ($i>85^{\circ}$) galaxies with a stellar mass within $\pm0.3$~dex of the Milky Way. With mass being a dominant driver of galaxy evolution (Gallazzi et al. 2005), removing this variable is crucial to determine the remaining drivers of evolution. Disk galaxies in this mass range show a variety of morphokinematic and star formation properties (Fisher \& Drory 2011), as well as diverse assembly histories (Somerville \& Dav{\'e} 2015). Aiming to represent that diversity, we choose a 2~dex range in SFR (see Fig. \ref{fig:geckos_sample}). This maximises the variety in assembly histories and increases the probability of detecting outflows, present in \mbox{$\sim30-50$\%} of main sequence galaxies (Stuber et al. 2021) and ubiquitous in starbursts (Veilleux et al. 2020). We avoid galaxies in the densest parts of clusters where external gas-removal processes are known to dominate disk evolution (e.g. Cortese et al. 2021). 

\begin{figure}[th]
    \includegraphics[width=\textwidth]{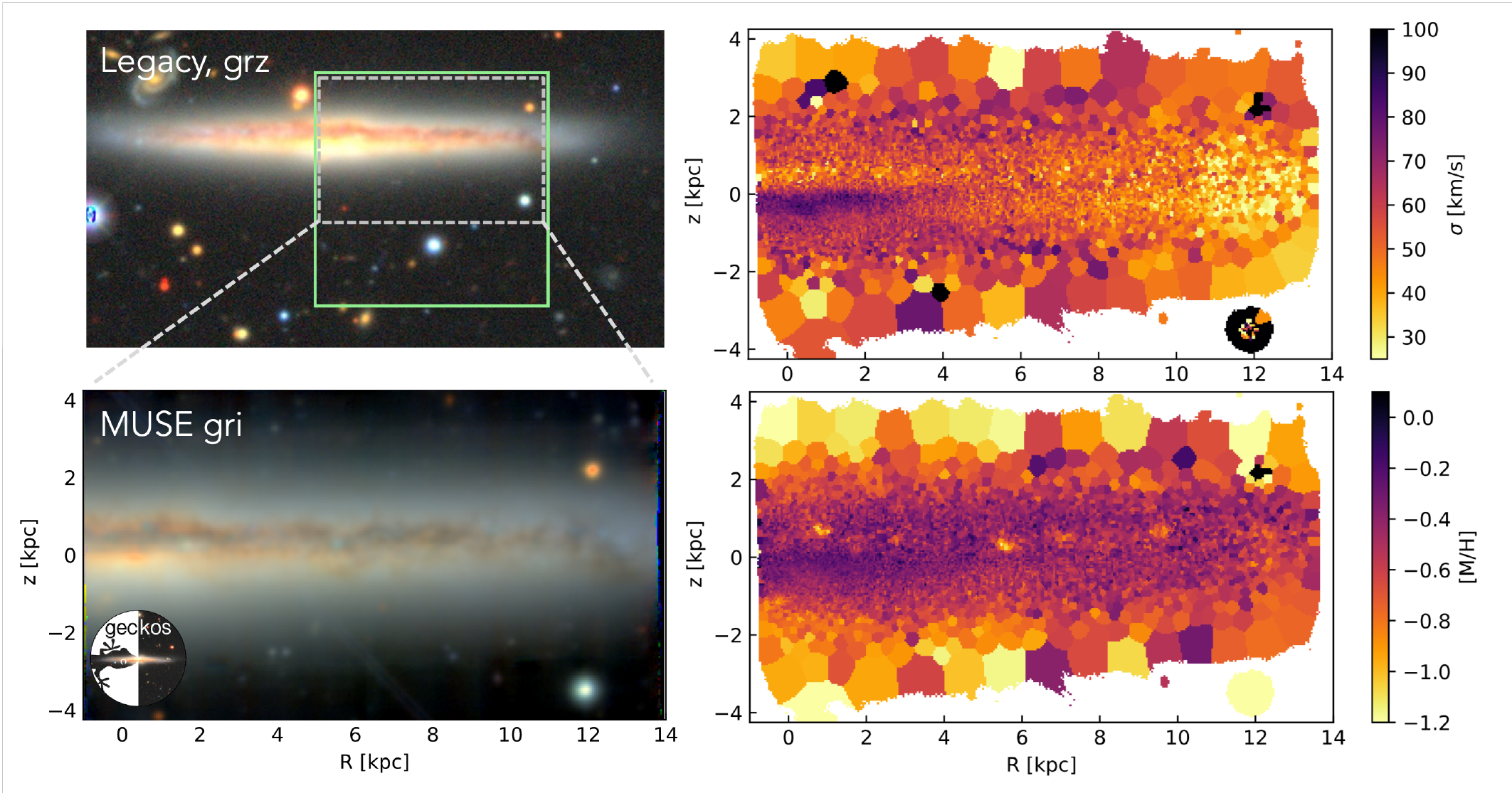}
    \caption{Preliminary analysis of GECKOS galaxy ESO120-016. On the top left, we show a DESI Legacy $grz$ colour image, with the MUSE field-of-view in green, whereas the bottom-left shows a reconstructed $gri$ colour image from the MUSE spectral data. Panels on the right show the stellar velocity dispersion $\sigma$ (top) and the mean stellar metallicity [M/H] (bottom), derived from spectra with wavelength range $\lambda=4750-7100$\AA\ using GIST combined with the MILES stellar population models from Vazdekis et al. (2015). The data have been Voronoi binned to obtain a constant spatial signal-to-noise of 40\AA$^{-1}$ per bin.} \label{fig:geckos_prelim}
\end{figure}

\vspace{0.3cm}
\noindent {\bf Observing Strategy:} Achieving the GECKOS science goals requires deep observations at large radius where the imprints of minor accretion events are most pronounced (Martig et al. 2021). The recovery of star formation histories using full-spectrum fitting requires \textbf{$S/N>40$} (Gallazzi et al. 2005, Walcher et al. 2009). To directly compare with chemodynamical models of the Milky Way in the solar environment, we aim to reach this $S/N$ target at a surface brightness of 23.5~V~mag~arcsec$^{-2}$ in the midplane (Melchior et al. 2007). For galaxies with a higher likelihood of extended outflows (0.3~dex above main-sequence) we ensure that the MUSE field-of-view reaches a height of 10~kpc, the known extent of the M82 outflow (Shopbell \& Bland-Hawthorn 1998). Outflow pointings are designed to reach 5$\times 10^{-18}$~erg~s~cm$^{-2}$ in a $\sim$0.5~kpc region (guided by observations of H$\alpha$ in M82). In Fig. \ref{fig:geckos_prelim} we present our preliminary results for ESO120-016, the first completed GECKOS galaxy (440min exposure time), where a clear radial stellar velocity dispersion and vertical metallicity gradient are visible.

\vspace{0.3cm}
\noindent {\bf Connecting to Simulations:} The GECKOS sample will be complemented by hydrodynamical simulations of galaxies to make predictions about the relative importance of the physical processes in disk galaxies like the Milky Way. For example, using the large-volume cosmological hydrodynamical EAGLE simulation, Mackereth et al. (2018) find that chemical bimodalities in Milky Way analogues are rare, while Evans et al. (2020) predict that only 5 per cent of Milky Way-mass galaxies undergo an early ($\sim$10 Gyrs ago), major accretion event. In contrast, using a small sample of galaxies from the NIHAO zoom-in simulations, Buck (2020) show that the $\alpha$-bimodality is a generic feature that originates from a gas-rich merger, but is not specific to an early major merger (see also Khoperskov et al. 2021).  

Furthermore, some high-resolution cosmological zoom-in simulations predict that mergers leave a clear kinematic and chemical signature in edge-on disks 
(Martig et al. 2014, Buck 2022, Garc{\'\i}a de la Cruz et al. 2021). This appears to be in conflict with the results from Yu et al. (2022) who show that while mergers onto the disk and other secular processes do affect the kinematics of simulated galaxies, they only play a secondary role in populating the thick disk and in-situ spheroid populations at redshift zero. Instead, they argue that stars with disk-like or spheroid-like orbits today were born that way.

By utilising GECKOS observations in combination with predictions from simulations, we will be able to study, among other things, outflow launching mechanisms, the efficiency of radial migration, and the connection between assembly history to the present-day structure of galaxies.
\vspace{0.3cm}

\noindent {\bf Survey Status:}
\noindent GECKOS VLT/MUSE observations began in October 2022, with expected completion by October 2024. The data will be reduced using \textsc{pymusepipe} (\url{https://github.com/emsellem/pymusepipe}) following the approach as outlined in Emsellem et al. (2022). For extracting the key measurables, e.g.  the stellar and gas kinematics, stellar population parameters, star formation histories, emission line fluxes, etc., we will employ a modified version of GIST (Bittner et al. 2019a,b) combined with other smaller analysis packages. The GECKOS team will deliver fully reduced data cubes, and 2D maps of all key measurables through the Australian data server Data Central (\url{https://datacentral.org.au/}), which will be used in combination with ESO phase 3 to release our data.

\section{Conclusion}
\noindent We present GECKOS, the first IFS survey to specifically target edge-on ($i > 85^{\circ}$) galaxies with a stellar mass within $\pm0.3$ dex of the Milky Way, uniquely complementing other MUSE galaxy surveys (e.g.~PHANGS, TIMER, MAD). GECKOS does not attempt to repeat previous multi-object surveys (e.g. CALIFA, MaNGA, SAMI) with higher resolution and signal-to-noise, but instead specifically addresses those questions that remain unanswered after years of large IFS programs. The survey is designed to concentrate on science above the midplane of galaxy disks that allows us to detect stars formed in the ancient thick disk as well as stars accreted from mergers. The main aim of GECKOS is to identify and constrain the fundamental processes that govern the formation and evolution of galactic disks.

\section*{Acknowledgements}
\noindent It is a great pleasure to thank the organisers and attendees for an enjoyable conference, and to our colleagues in the GECKOS Survey team (\url{https://geckos-survey.org/team.html}) for making this work possible. GECKOS is based on observations collected at the European Organisation for Astronomical Research in the Southern Hemisphere under ESO program 1110.B-4019(A). We wish to thank the ESO staff, and in particular the staff at Paranal Observatory, for carrying out the GECKOS observations. Part of this research was conducted by the Australian Research Council Centre of Excellence for All Sky Astrophysics in 3 Dimensions (ASTRO 3D), through project number CE170100013. JvdS acknowledges support of an Australian Research Council Discovery Early Career Research Award (project number DE200100461) funded by the Australian Government.

\end{document}